\documentclass[aps,pra,twocolumn,superscriptaddress,nopacs,letterpaper,longbibliography]{revtex4-1}
\usepackage[utf8]{inputenc}
\usepackage{amsmath}
\usepackage{color,graphicx}
\usepackage[pdfstartview=FitH,colorlinks=true,linkcolor=blue,citecolor=blue,urlcolor=blue]{hyperref}
\usepackage{lmodern}
\usepackage{bm}
\usepackage[english]{babel}

\newcommand{\eF}{\varepsilon_{F}}

\newcommand{\kF}{k_{\textrm{F}}}

\newcommand{\abs}[1]{\lvert{#1}\rvert}

\newcommand{\ssec}[1]{\noindent\emph{#1}---}

\begin{document}
\title{Universal Aspects of Vortex Reconnections across the BCS-BEC Crossover}
\date{\today}

\author{Marek Tylutki}
\affiliation{Faculty of Physics, Warsaw University of Technology, Ulica Koszykowa 75, PL-00662 Warsaw, Poland}
\author{Gabriel Wlaz\l{}owski}\email{gabriel.wlazlowski@pw.edu.pl}
\affiliation{Faculty of Physics, Warsaw University of Technology, Ulica Koszykowa 75, PL-00662 Warsaw, Poland}
\affiliation{Department of Physics, University of Washington, Seattle, Washington 98195--1560, USA}

\begin{abstract}
Reconnecting vortices in a superfluid allow for the energy transfer between different length scales and its subsequent dissipation. The present picture assumes that the dynamics of a reconnection is driven mostly by the phase of the order parameter, and this statement can be justified in the case of Bose-Einstein Condensates (BECs), where vortices have a simple internal structure. Therefore, it is natural to postulate that the reconnection dynamics in the vicinity of the reconnection moment is universal. This expectation has been confirmed in numerical simulations for BECs and experimentally for the superfluid ${}^4$He. Not much has been said about this relation in the context of Fermi superfluids. In this article we aim at bridging this gap, and we report our findings, which reveal that the reconnection dynamics conforms with the predicted universal behaviour across the entire BCS-BEC crossover. The universal scaling also survives for spin-imbalanced systems, where unpaired fermions induce a complex structure of the colliding vortices.
\end{abstract}

\maketitle

\ssec{Introduction}
\begin{figure}
\includegraphics[clip = true, width =.95\columnwidth]{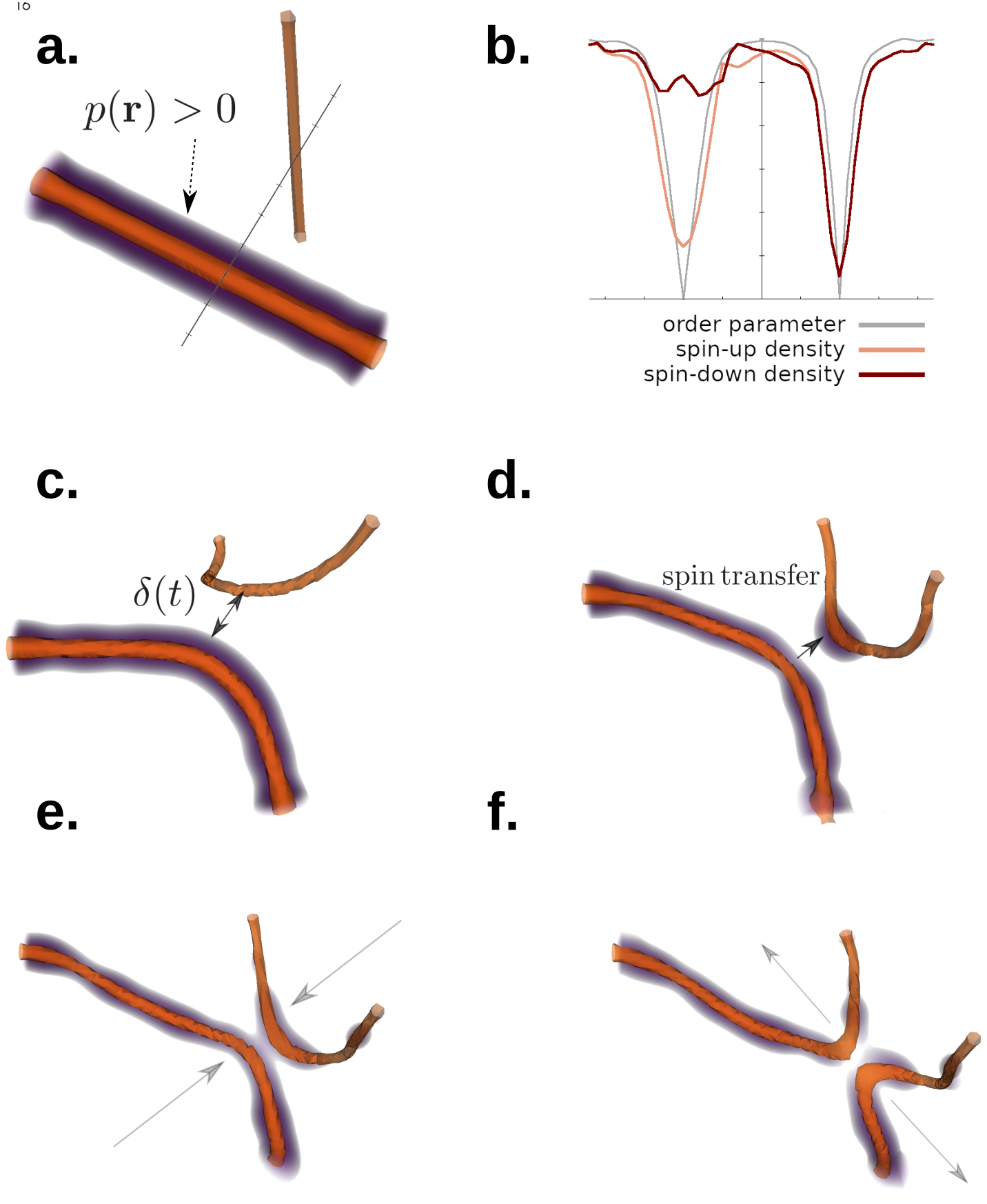}
\caption{Collision of two initially perpendicular vortex lines in strongly interacting Fermi gas. In this exotic case, one of the lines drags inside spin polarization $p(\bm{r})=\frac{n_{\uparrow}(\bm{r})-n_{\downarrow}(\bm{r})}{n_{\uparrow}(\bm{r})+n_{\downarrow}(\bm{r})}>0$. Profiles of basics characteristics (densities of spin components and order parameter) along line connecting the vortices is presented on panel b). Panels c)-f) provide selected snapshots from the simulation executed by means of time-dependent density functional theory. In the course of collision, the polarization tunnels through the superfluid and the vortices exchange the extra spin component. In this work we study scaling with time of the minimum distance between vortices $\delta(t)$ in fermionic superfluids.}
\label{fig.f0}
\end{figure}
\begin{figure*}[t]
\includegraphics[clip = true, width =.99\textwidth]{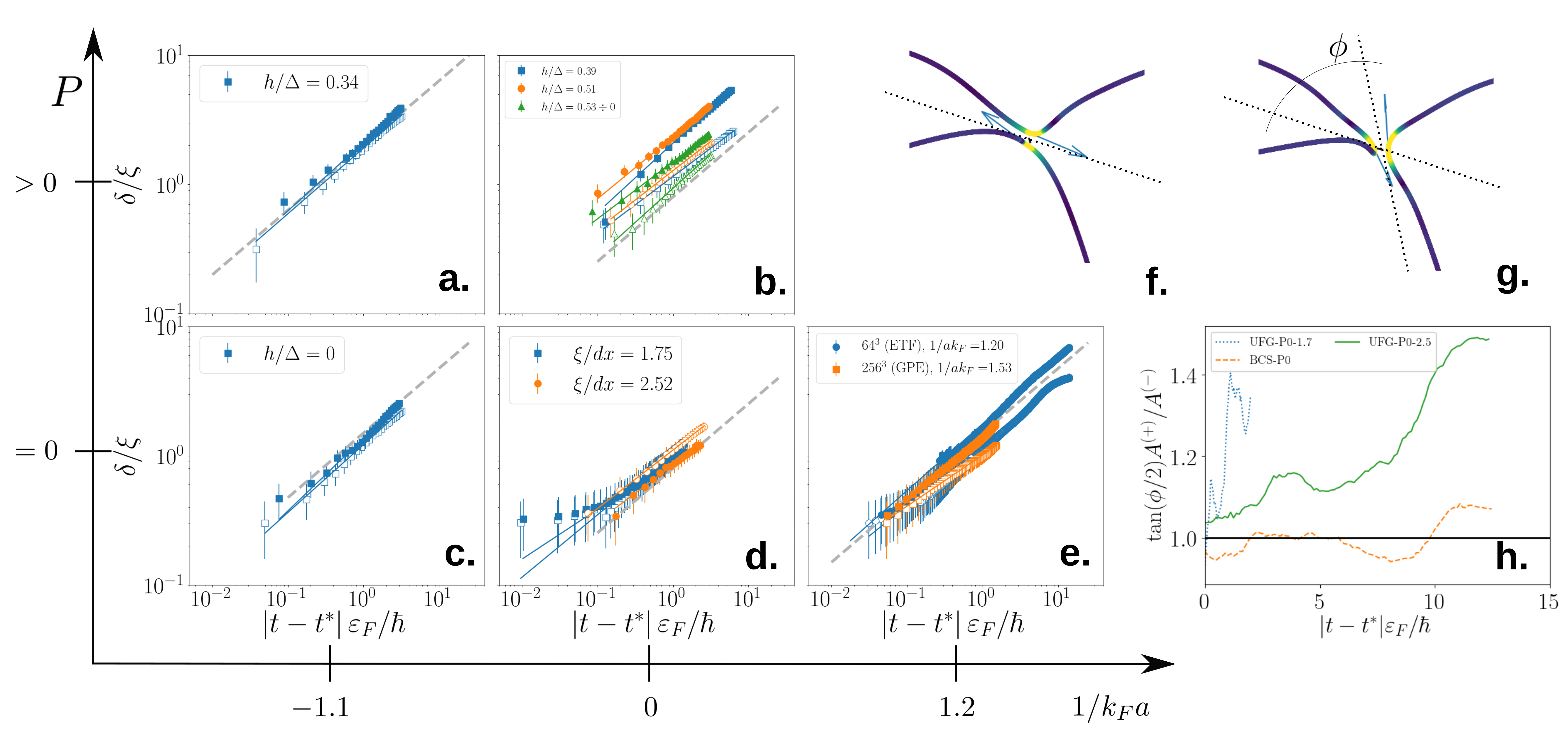}
\caption{Scaling of the minimal distance between the vortex lines as a function of time to the reconnection event. In panels a)-e) we show the minimal distance as a function of time to reconnection before (empty symbols) and after (full symbols) the collision for various polarizations and values of $1 / k_F a$. The solid lines show fits of the data to the relation in Eq.~\ref{eq.dist}. The fitted exponents are close to $1/2$ (marked by a dashed line) and have been measured as: (a) 0.51, 0.53; (b) 0.48, 0.47, 0.53, 0.51, 0.49, 0.48; (c) 0.52, 0.56; (d) 0.48, 0.47, 0.47, 0.41; and (e) 0.47, 0.52, 0.4, 0.5. For details see table~\ref{tab.t2}. In each panel, different data sets are in different colour (or grayscale) and with different symbols. In panels f) and g) we show the vortex configuration shortly before and shortly after the reconnection event, respectively for the unpolarized Fermi gas at unitarity. The color code (or grayscale) along the vortex line corresponds to its curvature and the arrows show tangent vectors to the vortex lines at the points of minimal distance. In panel h) we plot the expression $\tan (\phi / 2) A_{+} / A_{-}$, where $\phi$ is the angle between the asymptotes of the hyperbolae formed by the lines at the same time $t$ before and after the reconnection. We show three data sets, two for the UFG and one in the BCS regime. }
\label{fig.f1}
\end{figure*}
The quantum turbulent state is a tangle of vortex lines, and decays through their interactions, as was already proposed by Feynman~\cite{Feynman1955}. The reconnections of quantum vortex lines have thus been postulated as the principal mechanism of the decay of the quantum turbulent state~\cite{Paoletti2008,Henn2009,Tsatsos2016,Navon2019}, which made the understanding of reconnection dynamics crucial for this field. The numerical studies of reconnections were first performed in the context of the superfluid $^4$He with the vortex filament model~\cite{Schwarz1985,Schwarz1988} and then with the microscopic Gross–Pitaevskii equation (GPE)~\cite{Koplik1993}. The semiclassical model of vortex reconnection was proposed by Nazarenko and West~\cite{Nazarenko2003}, where the vortex lines before and after the reconnection were modelled as two hyperbolae on the same plane. The resulting solution implied that the relative distance $\delta$ between the lines scales with the time to the reconnection moment $t^*$ as $\sim |t-t^*|^{1/2}$. The same scaling could be less rigorously predicted by dimensional analysis assuming that the process of vortex collision is mainly driven by topology of the order parameter, as was first numerically verified for a BEC by A. Villois {\it et al.} in~\cite{Villois2017} and later also by Galantucci {\it et al.} in~\cite{Galantucci2019}. This assumption is justified for weakly interacting BECs, where we have one-to-one mapping between the order parameter and other characteristics such as velocity field or density distribution. In the recent Ref.~\cite{Galantucci2019} the extensive simulations of the GPE also test the scaling law at larger distances, where it changes to $\delta(t) \sim |t - t^*|$. Since the scaling is also in agreement with experimental measurements for superfluid helium~\cite{Fonda1924} it is natural to conjecture that the law should be universal. 

Despite many experiments with topological defects in ultracold gases of fermions~\cite{Zwierlein2005,Yefsah2013,Ku2014,Ku2016}, it has not been discussed whether or not the same scaling would be present in Fermi superfluids. These systems can be tuned from the BCS regime of weak attraction through the strongly interacting gas (unitarity limit) to the BEC regime of bosonic molecules, which is across the famous BCS-BEC crossover~\cite{Zwerger2012}. Vortices in a Fermi gas have a more complicated internal structure when compared to weakly interacting BECs~\cite{Machida2005,Sensarma2006}, and a direct relation between the order parameter and particle density or velocity field is lost. When introducing spin imbalance to the system, a degree of freedom that is not accessible for BECs, the vortex cores are filled by a surplus of the majority spin component~\cite{Takahashi2006,Hu2007}. Therefore, it is no longer obvious that the simple square-root scaling postulated by dimensional analysis for the BEC should be universal. In particular, it was recently observed that the vortex core in a polarized Fermi superfluid has a very rich structure due to the occupation of the Andreev states inside, and it can lead to the reverse circulation in the core~\cite{Magierski2020}. For larger spin polarization it is possible that the vortex is accompanied by the Fulde-Ferrell-Larkin-Ovchinnikov (FFLO) phase~\cite{Inotani2020} In Fig.~\ref{fig.f0}a we present configuration with two perpendicular vortices immersed in a strongly interacting Fermi gas. In the vicinity of one line (vertical), we applied an external magnetic field to induce spin polarization of the core $p(\bm{r}) = [n_{\uparrow}(\bm{r})-n_{\downarrow}(\bm{r})] / [n_{\uparrow}(\bm{r})+n_{\downarrow}(\bm{r})]$, where $n_{\sigma}$ stands for the particle density with spin $\sigma=\{\uparrow,\downarrow\}$. {\it A priori} it is not even clear if the vortex reconnection for such an exotic configuration would take place, but our direct numerical simulations (DNS) by means of density functional theory clearly demonstrate that vortices undergo a reconnection, as we present in Fig.~\ref{fig.f0}c-e and in the data shown in Fig. 2(b). Dynamics of the process is more complicated as compared to one observed in BECs: before the reconnection the vortices locally equilibrate via spin tunneling process (Fig.~\ref{fig.f0}c), and only then the exchange of lines takes place. In this article we test universality of the scaling law $\delta\sim |t-t^*|^{1/2}$ across the whole BCS-BEC crossover also allowing for effects arising from spin polarization. Due to the complexity of numerical simulations we restrict ourselves to testing the said scaling only for small distances between the vortex lines, where the square-root dependence is expected to hold on the BEC side. 

\ssec{The model}
A straightforward approach to accurately model the dynamics across the BCS-BEC crossover is to implement appropriate methods dedicated to each of the interaction regimes. It is convenient to parametrize the interaction strength via dimensionless quantity $1/a\kF$, where $a$ is an $s$-wave scattering length and $\kF=(3\pi^2n)^{1/3}$ is a Fermi wave vector corresponding to the total density $n$. The BCS side of the resonance ($1 / \kF a \lesssim -1$) corresponds to weak interactions where fermions with opposite spins form Cooper pairs. This is the regime of applicability of the mean-field Bogoliubov-de Gennes (BdG) equations, which define the evolution of quasiparticle wave functions $(u_\eta({\bf r},t),v_\eta({\bf r},t))^{T}$:
\begin{equation}
i\hbar\dfrac{\partial}{\partial t} \begin{pmatrix}
u_\eta({\bf r},t)\\
v_\eta({\bf r},t)\\
\end{pmatrix}
=
\mathcal{H}_{\rm BdG}
\begin{pmatrix}
u_\eta({\bf r},t)\\
v_\eta({\bf r},t)\\
\end{pmatrix},\label{eqn:tdbdg}
\end{equation}
with the Hamiltonian
\begin{equation}
\mathcal{H}_{\rm BdG} = \begin{pmatrix}
h_{\uparrow}({\bf r},t) - \mu_{\uparrow} & \Delta({\bf r},t) \\
\Delta^*({\bf r},t) & -h_{\downarrow}^{*}({\bf r},t) + \mu_{\downarrow} \\
\end{pmatrix}.\label{eqn:HBdG}
\end{equation}
Here $h_{\sigma}({\bf r},t)= -\hbar^2 \nabla^2 / 2m + V_{\sigma}({\bf r},t)$ is a single-quasiparticle Hamiltonian for spin channel $\sigma=\{\uparrow,\downarrow\}$ with external potential $V_{\sigma}$. $\Delta({\bf r},t) = -g\nu({\bf r},t)$
is a superfluid gap proportional to the anomalous density
\begin{equation*}
\nu({\bf r},t)= \dfrac{1}{2}\sum_{|E_\eta|<E_c} u_{\eta}({\bf r},t)v_{\eta}^{*}({\bf r},t)\left( f_{\beta}(-E_\eta)-f_{\beta}(E_\eta)\right),\label{eqn:nu}
\end{equation*}
and the coupling constant is $g=4\pi\hbar^2a/m$. $E_{\eta}$ denotes the quasiparticle energy and $E_c$ is energy cut-off scale required for regularization of the theory~\cite{Bulgac2002}. The Fermi distribution function $f_{\beta}(E)=1/(\exp(\beta E)+1)$ is introduced to model the temperature $k_B T=1/\beta$ effects and in the calculations we set $T \rightarrow 0$. Chemical potentials $\mu_{\sigma}$ are used to control the particle numbers when generating initial states for the time dependent calculations. These are obtained as self-consistent solutions of a static variant of Eqs.~(\ref{eqn:tdbdg}). The particle densities are computed as $n_{\uparrow}(\bm{r},t) = \sum_{|E_\eta|<E_c}\abs{u_{\eta}({\bf r},t)}^2 f_{\beta}(E_\eta)$ and $n_{\downarrow}(\bm{r},t) = \sum_{|E_\eta|<E_c}\abs{v_{\eta}({\bf r},t)}^2 f_{\beta}(-E_\eta)$.

On the other side of the resonance ($1 / \kF a \gtrsim +1$) the attraction is strong enough to bind atoms with opposite spins into spinless dimers, which subsequently condensate and form BEC: the Gross–Pitaevskii equation (GPE) becomes reliable in this regime. Here we used both the standard and the refined formulation of the GPE, known as the Extended Thomas-Fermi (ETF) model which take into account that dimers consist of two atoms of the fermionic type~\cite{Kim2004,Salasnich2009,Bulgac2014}:
\begin{equation*}
i \hbar \frac{\partial\Psi(\bm{r},t)}{\partial t} = \left( -\frac{\hbar^2}{4m}\nabla^2 + 2\frac{\delta\mathcal{E}_{\rm ETF}}{\delta n} + 2V_{\rm ext}(\bm{r},t) \right)\Psi(\bm{r},t),
\end{equation*}
and density is related to the condensate wave function as $n=2|\Psi|^2$. The extra factors of two appearing in the density definition, in the denominator of the kinetic term and in front of the interaction terms account for a composite structure of a dimer, which consist of two fermionic particles. The interaction term arises as a functional derivative of the density functional $\mathcal{E}_{\rm ETF} = \frac{3}{5} \eF n \,\xi (x + \xi) / [\xi + x (1 + \zeta) + 3 \pi \xi x^2] - \hbar^2 n / 2 m a^2$. The interaction parameter is $x = 1 / \kF a$, $\xi \approx 0.37$ is the Bertsch parameter and $\zeta = 0.901$ is a constant~\cite{Bulgac2014}. The choice of this form of the functional ensures that the energy of a uniform system scale as $E=\int \mathcal{E}_{\rm ETF}\,d^3\bm{r}\sim \frac{3}{5} \eF N$ for $0\lesssim x \lesssim 1$ (expected scaling for systems of fermionic nature), while in the deep BEC regime ($x\rightarrow\infty$) the energy scaling changes to $E\sim N^2$ as for the standard GPE.

Close to the resonance ($1 / \kF a \approx 0$) strong quantum correlations dominate and the system forms the unitary Fermi gas (UFG). Over the last decade, a formulation known as Asymmetric Superfluid Local Density Approximation (ASLDA) has demonstrated its validity in this strongly correlated regime~\cite{Bulgac2002,Bulgac2019,Wlazlowski2018}. The ASLDA approach utilizes concepts of density functional theory and overcomes deficiencies of the BdG approach when applied to strongly interacting systems. At a formal level time-dependent ASLDA equations have the same structure as Eqs~(\ref{eqn:tdbdg}) and (\ref{eqn:HBdG}), where the single-particle Hamiltonian and the pairing field are given by (in order to simplify the notation we omit the position and time dependence of the densities):
\begin{eqnarray*}
h_{\sigma}({\bf r},t)&=& -\frac{\hbar^2 \nabla^2}{2m^*} + U_{\sigma}(n_{\uparrow},n_{\downarrow},\nu) + V_{\sigma}({\bf r},t) ~,\\
\Delta({\bf r},t) & = & -\frac{\gamma}{(n_{\uparrow}+n_{\downarrow})^{1/3}}\nu ~.
\end{eqnarray*}
Here $U_{\sigma}(n_{\uparrow},n_{\downarrow},\nu)$ is a Hartree-Fock term which models the effects of strong interactions. Moreover, the pairing field $\Delta(n_{\uparrow},n_{\downarrow},\nu)$ depends on both particle densities and the anomalous density. In turn the ASLDA approach introduces additional coupling between the density modes and pairing modes, which is not present at the BdG level. Coupling constant $\gamma$ and further couplings entering the Hartree-Fock terms were tuned to provide a remarkable agreement with the quantum Monte Carlo calculations for both uniform and trapped systems at unitarity. For the explicit form of the single-particle Hamiltonian and the pairing field see section 9.3 in \cite{Zwerger2012}. In the calculations presented here we forced the effective mass $m^*$ to be equal the the bare mass $m$ in order to speed up our calculations. This modification does not induce any qualitative changes, and only quantitative modification at a level of a few percent is generated, see the Supplementary Material of~\cite{Bulgac2014}.

\ssec{Numerical procedure}
Direct numerical simulations with all three approaches were executed on a 3D spatial grid of size $64^3$ with a lattice spacing in the range $dx/\xi\in (1.75,2.5)$ when measured in units of coherence length $\xi$. We have tested that lattice of this size is sufficient to reproduce the scaling law $\delta \sim |t-t^*|^{1/2}$ for BEC regime with reasonable accuracy, in accordance with previous studies~\cite{Galantucci2019}. Tests with relatively low computationally demanding GPE also showed that increasing the grid size to $256^3$ does not change the results qualitatively. Studies are conducted in the zero temprature limit. Spatial derivatives were calculated using spectral methods, by means of Fourier transforms. Static BdG and ASLDA equations are solved by a series of successive diagonalizations of the Hamiltonian matrix~(\ref{eqn:HBdG}) until self-consistence. Static solution for the ETF model was generated by means of imaginary time projection method and for time dependent problems we used the Adams-Bashforth-Moulton (predictor-corrector) scheme of the fifth order as the integrator. More details on the implementation can be found in Refs.~\cite{Wlazlowski2018,GWCode}.

\begin{figure}[h]
\includegraphics[clip = true, width=0.99\columnwidth]{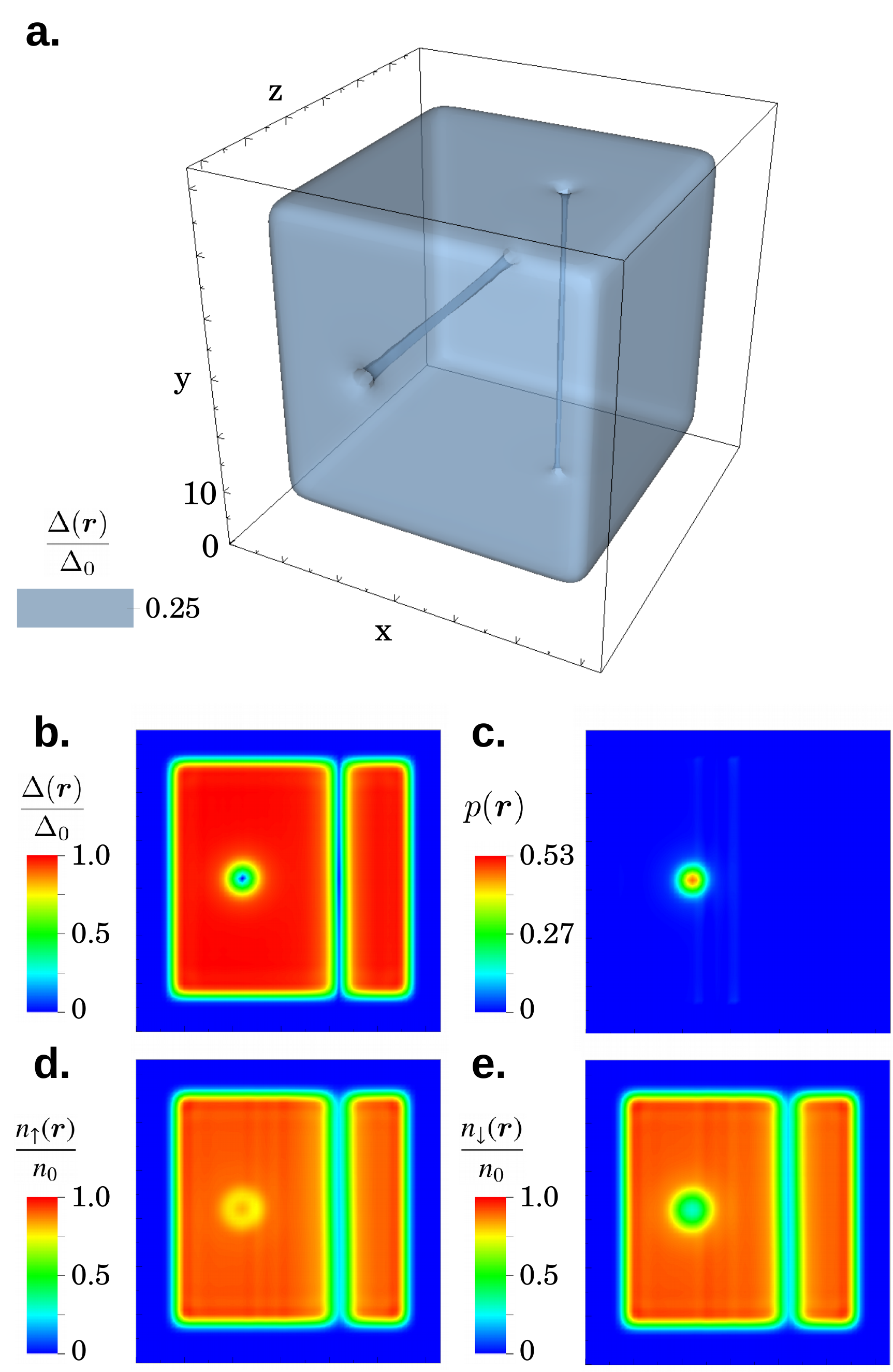}
\caption{Initial configuration of simulation presented in Fig.~\ref{fig.f0}: two perpendicular vortices in the unitary Fermi gas, where one of them is spin-polarized (data set UFG-P02). The contour plot (panel a.) shows a 3D visualization of the order parameter distribution for system being trapped in external potential~(\ref{eqn:Vext}). Panels b.-e. display cross section for $x-y$ plane of quantities: order parameter normalized to its bulk value $\Delta(\bm{r})/\Delta_{0}$ (b.), local polarization $p(\bm{r}) = [n_{\uparrow}(\bm{r})-n_{\downarrow}(\bm{r})] / [n_{\uparrow}(\bm{r})+n_{\downarrow}(\bm{r})]$ (c.), particle density of spin-up (d.) and spin-down (e.) components normalized to their bulk values.  } 
\label{fig:sm2}
\end{figure}

Lattice formulation imposes periodic boundary conditions. In  order to get rid of the interaction of vortices with their own images we introduced potential walls at the borders. We choose the potential to vary smoothly from zero in the bulk region to a finite value close to the boundary, so that all the derivatives are continuous. To this end we chose the potential at a single boundary of the form:
\begin{equation}
V_{\rm ext} (x) = V_0
\left\{\begin{array}{lll}
0 & \quad & x \leq l\\
s(x - l, w, \alpha) & \quad & l < x \leq l + w\\
1 & \quad & x \geq l + w\\
\end{array}\right.
\label{eqn:Vext}
\end{equation}
where $w$ is the width of the transient regime where $V_{\rm ext}$ has an intermediate value, and $l$ is the width measured from the box center where $V_{\rm ext}= 0$. The function $s(x, w, \alpha)$ has the form
\begin{equation}
s(x, w, \alpha) = \frac12 + \frac12 \tanh[\alpha \tan(\frac{\pi x}{w} - \frac{\pi}2)] ~.
\end{equation}
The total external potential is a sum of all such potential contributions for every wall of the simulation cube. In Fig.~\ref{fig:sm2} we show a 3D visualization of the simulation domain together with cross-sections demonstrating the spatial distribution of selected quantities, for the case corresponding to Fig.~\ref{fig.f0}. 

The protocol of our numerical experiments is the following: we start from initially straight, perpendicular vortex lines, separated by a distance $\delta_0$, in order not to favour either a parallel or antiparallel configuration. The starting point is obtained by imprinting the phase pattern of the order parameter (superfluid gap $\Delta$ for the Fermi superfluid or condensate's wave function $\Psi$ for the condensate of dimers) and subsequently allowing the system to find the lowest energy state corresponding to this topological constraint. In the case of a UFG or BCS regimes, this corresponds to iterating of the BdG and ASLDA equations until self-consistency is achieved, and in the case of the condensate of dimers it corresponds to the imaginary time evolution of the GPE. Next, we allow the system to evolve in time by executing time dependent variants of the methods. In the presented data sets, the vortex lines undergo a reconnection, which is preceded by the lines bending towards each other around the points where the reconnection will take place. After the reconnection the vortex lines separate and move apart, see visualization in Fig.~\ref{fig.f0}c-f. Finally, we extract the vortex lines from the data sets as a sequence of coordinates and analyze the dynamics of these lines. In particular, we measure the minimal distance $\delta(t)$ between the lines as they approach the reconnection and the tangent vectors of the vortex lines. The reconnection time $t^*$ has to be extrapolated, and we adopt two approaches: first we estimate the reconnection time as the weighted average of the time coordinates of the two points with the lowest distance $\delta$: $t^{*} = (\delta_{-} t_{+} + \delta_{+} t_{-}) / (\delta_{-} + \delta_{+})$, where $(-)$ refers to the earlier point in time and $(+)$ to the later point. Second, we estimate the same reconnection time by fitting the linear function $\delta(t)^2 = a_{\pm} t + b_{\pm}$ to both data sets before ($-$) and after ($+$) the provisionally chosen reconnection time, where we take only a restricted number of data points on each side. Then the reconnection time is estimated as: $t^* = -(b_{+} - b_{-}) / (a_{+} - a_{-})$. Both approaches provide estimates of $t^*$ wich are in good agreement.

\begin{figure}[h]
\includegraphics[clip = true, width=0.99\columnwidth]{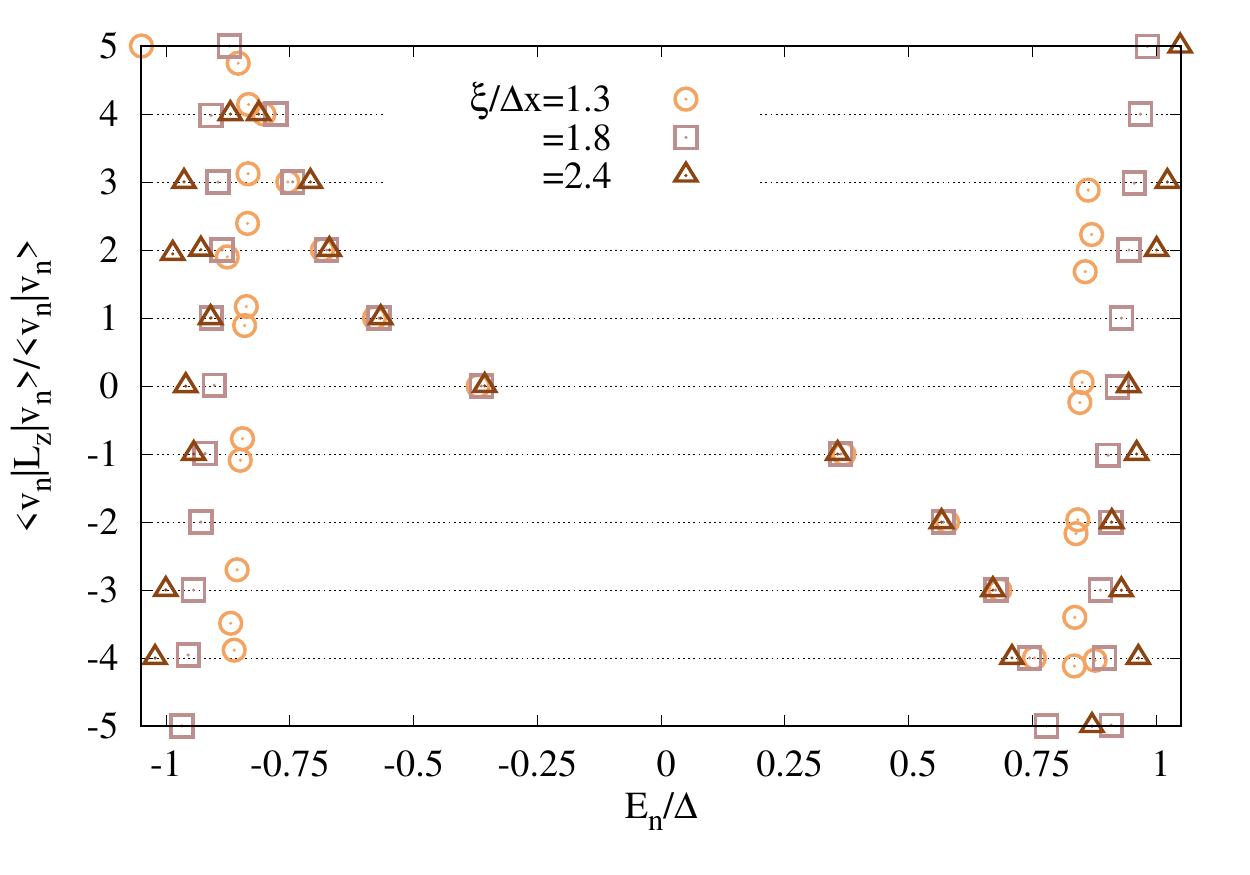}
\caption{Expectation values of the angular momentum of $\hat{L}_z$ for states residing in the paring gap $\Delta$ for vortex solution in a unitary Fermi gas (ASLDA method). Calculations were done for lattice of the same size as used for the reconnection studies ($64^3$). The resolution $\xi/dx\gtrsim 1.7$ is required to obtain correct numerical representation of the Andreev states. For lower lattice resolution, for example $\xi/\Delta x=1.3$, deviations from the integer value for the angular momentum are clearly visible. }
\label{fig:sm1}
\end{figure}

The solution representing a single straight vortex (along the $z$ direction) within BdG and ASLDA theory is expected to be:
\begin{equation}
\begin{pmatrix}
u_\eta({\bf r})\\
v_\eta({\bf r})\\
\end{pmatrix} = 
\begin{pmatrix}
u_\eta(r)e^{im\varphi}\\
v_\eta(r)e^{i(m+1)\varphi}\\
\end{pmatrix}e^{ik_z z},
\end{equation}
where $(r,\varphi,z)$ are cylindrical coordinates of point ${\bf r}$. Then, the quasiparticle wave functions have a well defined angular momentum $m=\frac{<v_\eta|\hat{L}_z|v_\eta>}{<v_\eta|v_\eta>}$. In the context of the investigation of the impact of the internal vortex structure on the reconnection dynamics it is important to have a correct representation of the states inside the vortex core. This is related to the question of a required lattice resolution. In Fig.~\ref{fig:sm1} we present numerically extracted values of the expectation values of the angular momentum of $\hat{L}_z$ for a vortex state solution with respect to the lattice resolution. The most important are states inside the pairing gap $\Delta$, called Andreev states, since they give rise to non-zero density inside vortices in fermionic systems. We find that for lattice resolution $\xi/dx\gtrsim 1.7$ the numerically represented wave functions have a well-defined angular momentum: the deviation of $\frac{<v_\eta|\hat{L}_z|v_\eta>}{<v_\eta|v_\eta>}$ from the nearest integer value is no more than 0.01. 

For extracting the positions of the vortices we used the algorithm presented in~\cite{Villois2016}, with the modification that instead of searching for zeros in the paring field/condensate wave function, we were searching for points around which its phase has a non-zero winding number. This modification greatly improves the stability of the tracking algorithm for the cases of BdG and ASLDA calculations. The algorithm searches for the vortex lines' positions with subgrid resolution and the precision of extracting the positions of individual lines was estimated to be $0.2dx$ and the accuracy of the measurement of relative distance is $\Delta\delta\approx 0.46\,dx$. 

\ssec{Results}
Numerical results for the minimal distance $\delta(t)$ are tested against their compatibility with the formula 
\begin{equation}
\delta(t) = A_{\pm} |t - t^*|^{\alpha_{\pm}}
\label{eq.dist}
\end{equation} where the fitting coefficients $A$ and $\alpha$ are allowed to be different before ($-$) and after ($+$) the reconnection. In Fig.~\ref{fig.f1}c-e we present numerically extracted values of $\delta(t)$ for systems being in the three regimes: BCS regime ($1 / k_F a = -1.1$), unitary regime ($1 / k_F a = 0$) and BEC regime ($1 / k_F a = +1.2$). In these runs we imposed the system to be spin-symmetric, i.e number of spin-up $N_{\uparrow}$ and spin-down $N_{\downarrow}$ particles are equal. For better visibility, the distance between the reconnecting points as a function of the time to the reconnection event is plotted in the logarithmic scale together with the slope corresponding to $\sim t^{1/2}$. For the BEC regime (panel e) the results for a lattice of size $64^3$ and $256^3$ are provided and demonstrate robustness of our results with respect to the lattice size.  In the plot referring to the unitary regime (panel d) we show the results for different temporal and spatial resolutions. Clearly, there is no qualitative change between these two data sets. In most cases, we observe that the separation of the vortex lines is faster than their approach before the collision, and this is reflected in the coefficient $A_{\pm}$ in front of the Eq.~(\ref{eq.dist}). This finding is consistent with the recent work, Ref.~\cite{PhysRevLett.125.164501}. The exponent in the power law arising from the fitting procedure remains close to $\alpha_{\pm} \approx 1/2$. Reduced chi-square statistics for deviations from the expected law $|t-t^*|^{1/2}$ stays below one for each case, confirming compatibility. The fitted exponents are summarized in the table~\ref{tab.t2}. In the supplemental material~\cite{supplemental} we provide a list of movies with the visualization for each case. 
\begin{table*}[t]
\centering
\begin{tabular}{cccccccccc}
data set label & $N_{\uparrow}$ & $N_{\downarrow}$ & $1 / \kF a$ & $\xi/dx$ & $\Delta t\eF/\hbar$ & $h/\Delta$& $p_{\textrm{core}}$ & $\alpha_{+}$ & $\alpha_{-}$ \\
\hline\hline
BCS-P0 & 1906 & 1906 & -1.1 & 2.52 & 0.125 & 0 & 0 & 0.52 & 0.56\\
UFG-P0-1.7  & 806 & 806 & 0 & 1.75 & 0.020 & 0 & 0 & 0.48 & 0.47 \\
UFG-P0-2.5 & 254 & 254 & 0 & 2.52 & 0.125 & 0 & 0 & 0.47 & 0.41 \\
BEC-64 & 800 & 800 & 1.20 & 2.02 & 0.007 & 0 & 0 & 0.47 & 0.52 \\
BEC-256 & 800 & 800 & 1.53 & 11.58 & 0.022 & 0 & 0 & 0.40 & 0.50 \\
\hline
BCS-P1  & 1943 & 1867 & -1.1 & 2.52 & 0.125 & 0.34 & 0.15 & 0.51 & 0.53 \\
UFG-P1  & 811 & 802 & 0 & 1.75 & 0.248 & 0.39 & 0.39 & 0.48 & 0.47 \\
UFG-P2  & 828 & 811 & 0 & 1.75 & 0.125 & 0.51 & 0.44 & 0.43 & 0.51 \\
UFG-P02  & 743 & 736 & 0 & 1.75 & 0.125 & 0.53$\div$0 & 0.47$\div$0 & 0.49 & 0.48 \\
\end{tabular}
\caption{Summary of simulations parameters. All calculations were executed on spatial lattice of size $64^3$,  except for the data set BEC-256 where the lattice size was increased to $256^3$. $N_\sigma$ denotes the number of particles of a given spin type, $1 / \kF a$ defines the interaction parameter, $\xi/dx$ defines the spatial lattice resolution while $\Delta t\eF/\hbar$ is a temporal resolution of the trajectories. In the case of simulations for spin-imbalanced systems we provide a mismatch between chemical potentials $h=(\mu_{\uparrow}-\mu_{\downarrow})/2$ with respect to the paring gap, and $p_{\textrm{core}}$ shows the value of the induced spin polarization inside the vortex core. The data set UFG-P02 shows the conditions for the simulation presented in Fig.~1, where only one vortex line was polarized. In the last two columns we provide values of the scaling law exponents emerging from the fitting procedure.}
\label{tab.t2}
\end{table*}

The tests are supplemented by runs for spin imbalanced systems $N_{\uparrow}>N_{\downarrow}$, see Figs~\ref{fig.f1}a-b, with the same conclusions as for the spin-symmetric cases. Also an exotic case, where only one vortex line drags the polarization has been considered, as shown on Fig.~\ref{fig.f0}, and it also exhibits postulated scaling. The imbalanced runs were realized by imposing difference of chemical potentials between spin-components $h=(\mu_{\uparrow}-\mu_{\downarrow})/2$ in BdG and ASLDA equations. The difference is kept below the Chandrasekhar-Clogston limit $h < \Delta$ (where $\Delta$ stands for the superfluid gap) in order not to destroy the superfluidity in the bulk and, at the same time, it is large enough to generate significant polarization $p$ of the vortex cores. 

Finally, we are interested in the geometry of colliding vortex lines, and to this end we calculate the tangent vectors along the lines. Motivated by the theory presented in Ref.~\cite{Villois2017} and simulations for the BEC therein, we want to check if the reconnection is always between the locally antiparallel segments of the vortex lines, and to test the relation $\tan \phi / 2 = A_{-} / A_{+}$ (assuming the exponent $\alpha_\pm = 1/2$ is the same before and after the reconnection) for the BCS-BEC crossover. Here $\phi$ is the angle between the asymptotes of the hyperbolae formed by vortex lines before and after the collision, see Figs~\ref{fig.f1}f-g. In panel (h) the expression $A_{+} \tan (\phi / 2) / A_{-} $, which is expected to be one for the reconnection, is plotted against the time to the reconnection. We confirm that the values are approaching one as the time to the reconnection decreases. 

\ssec{Conclusions}
In conclusion, we extended the analysis of the reconneting vortex lines to the wide range of physical systems and observe a similar behaviour as was predicted, simulated and observed for the weakly interacting BECs. As a result of the analysis, the scaling of the distance between the reconnection points of the two lines as a function of time in the form $\delta(t)\sim |t-t^*|^{1/2}$ can be regarded as universal and conforms to predictions from simplified theoretical models. This universality can be attributed to the mechanism of the reconnection which is primarily driven by the topology (phase) of these defects, which is universal for all superfluids. 

\section*{Acknowledgements}
We would like to thank Piotr Magierski for useful discussions. This work was supported by the Polish National Science Center (NCN), Contract No. UMO-2017/26/E/ST3/00428. We acknowledge PRACE for awarding us access to Piz Daint based in Switzerland at Swiss National Supercomputing Centre (CSCS), decision No. 2019215113. We also acknowledge  Interdisciplinary Centre for Mathematical and Computational Modelling (ICM) of Warsaw University for providing us with computational resources
(grant No. GA83-9).

%


\begin{thebibliography}{31}%
\makeatletter
\providecommand \@ifxundefined [1]{%
 \@ifx{#1\undefined}
}%
\providecommand \@ifnum [1]{%
 \ifnum #1\expandafter \@firstoftwo
 \else \expandafter \@secondoftwo
 \fi
}%
\providecommand \@ifx [1]{%
 \ifx #1\expandafter \@firstoftwo
 \else \expandafter \@secondoftwo
 \fi
}%
\providecommand \natexlab [1]{#1}%
\providecommand \enquote  [1]{``#1''}%
\providecommand \bibnamefont  [1]{#1}%
\providecommand \bibfnamefont [1]{#1}%
\providecommand \citenamefont [1]{#1}%
\providecommand \href@noop [0]{\@secondoftwo}%
\providecommand \href [0]{\begingroup \@sanitize@url \@href}%
\providecommand \@href[1]{\@@startlink{#1}\@@href}%
\providecommand \@@href[1]{\endgroup#1\@@endlink}%
\providecommand \@sanitize@url [0]{\catcode `\\12\catcode `\$12\catcode
  `\&12\catcode `\#12\catcode `\^12\catcode `\_12\catcode `\%12\relax}%
\providecommand \@@startlink[1]{}%
\providecommand \@@endlink[0]{}%
\providecommand \url  [0]{\begingroup\@sanitize@url \@url }%
\providecommand \@url [1]{\endgroup\@href {#1}{\urlprefix }}%
\providecommand \urlprefix  [0]{URL }%
\providecommand \Eprint [0]{\href }%
\providecommand \doibase [0]{http://dx.doi.org/}%
\providecommand \selectlanguage [0]{\@gobble}%
\providecommand \bibinfo  [0]{\@secondoftwo}%
\providecommand \bibfield  [0]{\@secondoftwo}%
\providecommand \translation [1]{[#1]}%
\providecommand \BibitemOpen [0]{}%
\providecommand \bibitemStop [0]{}%
\providecommand \bibitemNoStop [0]{.\EOS\space}%
\providecommand \EOS [0]{\spacefactor3000\relax}%
\providecommand \BibitemShut  [1]{\csname bibitem#1\endcsname}%
\let\auto@bib@innerbib\@empty
\bibitem [{\citenamefont {Feynman}(1955)}]{Feynman1955}%
  \BibitemOpen
  \bibfield  {author} {\bibinfo {author} {\bibfnamefont {R.P.}\ \bibnamefont
  {Feynman}},\ }\bibfield  {title} {\enquote {\bibinfo {title} {Chapter ii
  application of quantum mechanics to liquid helium},}\ \ }(\bibinfo
  {publisher} {Elsevier},\ \bibinfo {year} {1955})\ pp.\ \bibinfo {pages} {17
  -- 53}\BibitemShut {NoStop}%
\bibitem [{\citenamefont {Paoletti}\ \emph {et~al.}(2008)\citenamefont
  {Paoletti}, \citenamefont {Fisher}, \citenamefont {Sreenivasan},\ and\
  \citenamefont {Lathrop}}]{Paoletti2008}%
  \BibitemOpen
  \bibfield  {author} {\bibinfo {author} {\bibfnamefont {M.~S.}\ \bibnamefont
  {Paoletti}}, \bibinfo {author} {\bibfnamefont {Michael~E.}\ \bibnamefont
  {Fisher}}, \bibinfo {author} {\bibfnamefont {K.~R.}\ \bibnamefont
  {Sreenivasan}}, \ and\ \bibinfo {author} {\bibfnamefont {D.~P.}\ \bibnamefont
  {Lathrop}},\ }\bibfield  {title} {\enquote {\bibinfo {title} {Velocity
  {Statistics} {Distinguish} {Quantum} {Turbulence} from {Classical}
  {Turbulence}},}\ }\href {\doibase 10.1103/PhysRevLett.101.154501} {\bibfield
  {journal} {\bibinfo  {journal} {Phys. Rev. Lett.}\ }\textbf {\bibinfo
  {volume} {101}},\ \bibinfo {pages} {154501} (\bibinfo {year}
  {2008})}\BibitemShut {NoStop}%
\bibitem [{\citenamefont {Henn}\ \emph {et~al.}(2009)\citenamefont {Henn},
  \citenamefont {Seman}, \citenamefont {Roati}, \citenamefont {Magalh\~aes},\
  and\ \citenamefont {Bagnato}}]{Henn2009}%
  \BibitemOpen
  \bibfield  {author} {\bibinfo {author} {\bibfnamefont {E.~A.~L.}\
  \bibnamefont {Henn}}, \bibinfo {author} {\bibfnamefont {J.~A.}\ \bibnamefont
  {Seman}}, \bibinfo {author} {\bibfnamefont {G.}~\bibnamefont {Roati}},
  \bibinfo {author} {\bibfnamefont {K.~M.~F.}\ \bibnamefont {Magalh\~aes}}, \
  and\ \bibinfo {author} {\bibfnamefont {V.~S.}\ \bibnamefont {Bagnato}},\
  }\bibfield  {title} {\enquote {\bibinfo {title} {Emergence of turbulence in
  an oscillating bose-einstein condensate},}\ }\href {\doibase
  10.1103/PhysRevLett.103.045301} {\bibfield  {journal} {\bibinfo  {journal}
  {Phys. Rev. Lett.}\ }\textbf {\bibinfo {volume} {103}},\ \bibinfo {pages}
  {045301} (\bibinfo {year} {2009})}\BibitemShut {NoStop}%
\bibitem [{\citenamefont {Tsatsos}\ \emph {et~al.}(2016)\citenamefont
  {Tsatsos}, \citenamefont {Tavares}, \citenamefont {Cidrim}, \citenamefont
  {Fritsch}, \citenamefont {Caracanhas}, \citenamefont {{dos Santos}},
  \citenamefont {Barenghi},\ and\ \citenamefont {Bagnato}}]{Tsatsos2016}%
  \BibitemOpen
  \bibfield  {author} {\bibinfo {author} {\bibfnamefont {Marios~C.}\
  \bibnamefont {Tsatsos}}, \bibinfo {author} {\bibfnamefont {Pedro~E.S.}\
  \bibnamefont {Tavares}}, \bibinfo {author} {\bibfnamefont {André}\
  \bibnamefont {Cidrim}}, \bibinfo {author} {\bibfnamefont {Amilson~R.}\
  \bibnamefont {Fritsch}}, \bibinfo {author} {\bibfnamefont {Mônica~A.}\
  \bibnamefont {Caracanhas}}, \bibinfo {author} {\bibfnamefont
  {F.~Ednilson~A.}\ \bibnamefont {{dos Santos}}}, \bibinfo {author}
  {\bibfnamefont {Carlo~F.}\ \bibnamefont {Barenghi}}, \ and\ \bibinfo {author}
  {\bibfnamefont {Vanderlei~S.}\ \bibnamefont {Bagnato}},\ }\bibfield  {title}
  {\enquote {\bibinfo {title} {Quantum turbulence in trapped atomic
  bose–einstein condensates},}\ }\href {\doibase
  https://doi.org/10.1016/j.physrep.2016.02.003} {\bibfield  {journal}
  {\bibinfo  {journal} {Physics Reports}\ }\textbf {\bibinfo {volume} {622}},\
  \bibinfo {pages} {1 -- 52} (\bibinfo {year} {2016})},\ \bibinfo {note}
  {quantum turbulence in trapped atomic Bose–Einstein
  condensates}\BibitemShut {NoStop}%
\bibitem [{\citenamefont {Navon}\ \emph {et~al.}(2019)\citenamefont {Navon},
  \citenamefont {Eigen}, \citenamefont {Zhang}, \citenamefont {Lopes},
  \citenamefont {Gaunt}, \citenamefont {Fujimoto}, \citenamefont {Tsubota},
  \citenamefont {Smith},\ and\ \citenamefont {Hadzibabic}}]{Navon2019}%
  \BibitemOpen
  \bibfield  {author} {\bibinfo {author} {\bibfnamefont {Nir}\ \bibnamefont
  {Navon}}, \bibinfo {author} {\bibfnamefont {Christoph}\ \bibnamefont
  {Eigen}}, \bibinfo {author} {\bibfnamefont {Jinyi}\ \bibnamefont {Zhang}},
  \bibinfo {author} {\bibfnamefont {Raphael}\ \bibnamefont {Lopes}}, \bibinfo
  {author} {\bibfnamefont {Alexander~L.}\ \bibnamefont {Gaunt}}, \bibinfo
  {author} {\bibfnamefont {Kazuya}\ \bibnamefont {Fujimoto}}, \bibinfo {author}
  {\bibfnamefont {Makoto}\ \bibnamefont {Tsubota}}, \bibinfo {author}
  {\bibfnamefont {Robert~P.}\ \bibnamefont {Smith}}, \ and\ \bibinfo {author}
  {\bibfnamefont {Zoran}\ \bibnamefont {Hadzibabic}},\ }\bibfield  {title}
  {\enquote {\bibinfo {title} {Synthetic dissipation and cascade fluxes in a
  turbulent quantum gas},}\ }\href {\doibase 10.1126/science.aau6103}
  {\bibfield  {journal} {\bibinfo  {journal} {Science}\ }\textbf {\bibinfo
  {volume} {366}},\ \bibinfo {pages} {382--385} (\bibinfo {year}
  {2019})}\BibitemShut {NoStop}%
\bibitem [{\citenamefont {Schwarz}(1985)}]{Schwarz1985}%
  \BibitemOpen
  \bibfield  {author} {\bibinfo {author} {\bibfnamefont {K.~W.}\ \bibnamefont
  {Schwarz}},\ }\bibfield  {title} {\enquote {\bibinfo {title}
  {Three-dimensional vortex dynamics in superfluid {He} 4 : {Line}-line and
  line-boundary interactions},}\ }\href {\doibase 10.1103/PhysRevB.31.5782}
  {\bibfield  {journal} {\bibinfo  {journal} {Phys. Rev. B}\ }\textbf {\bibinfo
  {volume} {31}},\ \bibinfo {pages} {5782--5804} (\bibinfo {year}
  {1985})}\BibitemShut {NoStop}%
\bibitem [{\citenamefont {Schwarz}(1988)}]{Schwarz1988}%
  \BibitemOpen
  \bibfield  {author} {\bibinfo {author} {\bibfnamefont {K.~W.}\ \bibnamefont
  {Schwarz}},\ }\bibfield  {title} {\enquote {\bibinfo {title}
  {Three-dimensional vortex dynamics in superfluid {He} 4 : {Homogeneous}
  superfluid turbulence},}\ }\href {\doibase 10.1103/PhysRevB.38.2398}
  {\bibfield  {journal} {\bibinfo  {journal} {Phys. Rev. B}\ }\textbf {\bibinfo
  {volume} {38}},\ \bibinfo {pages} {2398--2417} (\bibinfo {year}
  {1988})}\BibitemShut {NoStop}%
\bibitem [{\citenamefont {Koplik}\ and\ \citenamefont
  {Levine}(1993)}]{Koplik1993}%
  \BibitemOpen
  \bibfield  {author} {\bibinfo {author} {\bibfnamefont {Joel}\ \bibnamefont
  {Koplik}}\ and\ \bibinfo {author} {\bibfnamefont {Herbert}\ \bibnamefont
  {Levine}},\ }\bibfield  {title} {\enquote {\bibinfo {title} {Vortex
  reconnection in superfluid helium},}\ }\href {\doibase
  10.1103/PhysRevLett.71.1375} {\bibfield  {journal} {\bibinfo  {journal}
  {Phys. Rev. Lett.}\ }\textbf {\bibinfo {volume} {71}},\ \bibinfo {pages}
  {1375--1378} (\bibinfo {year} {1993})}\BibitemShut {NoStop}%
\bibitem [{\citenamefont {Nazarenko}\ and\ \citenamefont
  {West}(2003)}]{Nazarenko2003}%
  \BibitemOpen
  \bibfield  {author} {\bibinfo {author} {\bibfnamefont {Sergey}\ \bibnamefont
  {Nazarenko}}\ and\ \bibinfo {author} {\bibfnamefont {Robert}\ \bibnamefont
  {West}},\ }\bibfield  {title} {\enquote {\bibinfo {title} {Analytical
  solution for nonlinear schrödinger vortex reconnection},}\ }\href {\doibase
  doi.org/10.1023/A:1023719007403} {\bibfield  {journal} {\bibinfo  {journal}
  {J Low Temp Phys}\ }\textbf {\bibinfo {volume} {132}},\ \bibinfo {pages} {1
  -- 10} (\bibinfo {year} {2003})}\BibitemShut {NoStop}%
\bibitem [{\citenamefont {Villois}\ \emph {et~al.}(2017)\citenamefont
  {Villois}, \citenamefont {Proment},\ and\ \citenamefont
  {Krstulovic}}]{Villois2017}%
  \BibitemOpen
  \bibfield  {author} {\bibinfo {author} {\bibfnamefont {Alberto}\ \bibnamefont
  {Villois}}, \bibinfo {author} {\bibfnamefont {Davide}\ \bibnamefont
  {Proment}}, \ and\ \bibinfo {author} {\bibfnamefont {Giorgio}\ \bibnamefont
  {Krstulovic}},\ }\bibfield  {title} {\enquote {\bibinfo {title} {Universal
  and nonuniversal aspects of vortex reconnections in superfluids},}\ }\href
  {\doibase 10.1103/PhysRevFluids.2.044701} {\bibfield  {journal} {\bibinfo
  {journal} {Phys. Rev. Fluids}\ }\textbf {\bibinfo {volume} {2}},\ \bibinfo
  {pages} {044701} (\bibinfo {year} {2017})}\BibitemShut {NoStop}%
\bibitem [{\citenamefont {Galantucci}\ \emph {et~al.}(2019)\citenamefont
  {Galantucci}, \citenamefont {Baggaley}, \citenamefont {Parker},\ and\
  \citenamefont {Barenghi}}]{Galantucci2019}%
  \BibitemOpen
  \bibfield  {author} {\bibinfo {author} {\bibfnamefont {Luca}\ \bibnamefont
  {Galantucci}}, \bibinfo {author} {\bibfnamefont {Andrew~W.}\ \bibnamefont
  {Baggaley}}, \bibinfo {author} {\bibfnamefont {Nick~G.}\ \bibnamefont
  {Parker}}, \ and\ \bibinfo {author} {\bibfnamefont {Carlo~F.}\ \bibnamefont
  {Barenghi}},\ }\bibfield  {title} {\enquote {\bibinfo {title} {Crossover from
  interaction to driven regimes in quantum vortex reconnections},}\ }\href
  {\doibase 10.1073/pnas.1818668116} {\bibfield  {journal} {\bibinfo  {journal}
  {Proc Natl Acad Sci USA}\ }\textbf {\bibinfo {volume} {116}},\ \bibinfo
  {pages} {12204--12211} (\bibinfo {year} {2019})}\BibitemShut {NoStop}%
\bibitem [{\citenamefont {Fonda}\ \emph {et~al.}(2019)\citenamefont {Fonda},
  \citenamefont {Sreenivasan},\ and\ \citenamefont {Lathrop}}]{Fonda1924}%
  \BibitemOpen
  \bibfield  {author} {\bibinfo {author} {\bibfnamefont {Enrico}\ \bibnamefont
  {Fonda}}, \bibinfo {author} {\bibfnamefont {Katepalli~R}\ \bibnamefont
  {Sreenivasan}}, \ and\ \bibinfo {author} {\bibfnamefont {Daniel~P}\
  \bibnamefont {Lathrop}},\ }\bibfield  {title} {\enquote {\bibinfo {title}
  {{Reconnection scaling in quantum fluids}},}\ }\href {\doibase
  10.1073/pnas.1816403116} {\bibfield  {journal} {\bibinfo  {journal}
  {Proceedings of the National Academy of Sciences}\ }\textbf {\bibinfo
  {volume} {116}},\ \bibinfo {pages} {1924--1928} (\bibinfo {year}
  {2019})}\BibitemShut {NoStop}%
\bibitem [{\citenamefont {Zwierlein}\ \emph {et~al.}(2005)\citenamefont
  {Zwierlein}, \citenamefont {Abo-Shaeer}, \citenamefont {Schirotzek},
  \citenamefont {Schunck},\ and\ \citenamefont {Ketterle}}]{Zwierlein2005}%
  \BibitemOpen
  \bibfield  {author} {\bibinfo {author} {\bibfnamefont {M.~W.}\ \bibnamefont
  {Zwierlein}}, \bibinfo {author} {\bibfnamefont {J.~R.}\ \bibnamefont
  {Abo-Shaeer}}, \bibinfo {author} {\bibfnamefont {A.}~\bibnamefont
  {Schirotzek}}, \bibinfo {author} {\bibfnamefont {C.~H.}\ \bibnamefont
  {Schunck}}, \ and\ \bibinfo {author} {\bibfnamefont {W.}~\bibnamefont
  {Ketterle}},\ }\bibfield  {title} {\enquote {\bibinfo {title} {Vortices and
  superfluidity in a strongly interacting fermi gas},}\ }\href {\doibase
  10.1038/nature03858} {\bibfield  {journal} {\bibinfo  {journal} {Nature}\
  }\textbf {\bibinfo {volume} {435}},\ \bibinfo {pages} {1047--1051} (\bibinfo
  {year} {2005})}\BibitemShut {NoStop}%
\bibitem [{\citenamefont {Yefsah}\ \emph {et~al.}(2013)\citenamefont {Yefsah},
  \citenamefont {Sommer}, \citenamefont {Ku}, \citenamefont {Cheuk},
  \citenamefont {Ji}, \citenamefont {Bakr},\ and\ \citenamefont
  {Zwierlein}}]{Yefsah2013}%
  \BibitemOpen
  \bibfield  {author} {\bibinfo {author} {\bibfnamefont {Tarik}\ \bibnamefont
  {Yefsah}}, \bibinfo {author} {\bibfnamefont {Ariel~T.}\ \bibnamefont
  {Sommer}}, \bibinfo {author} {\bibfnamefont {Mark J.~H.}\ \bibnamefont {Ku}},
  \bibinfo {author} {\bibfnamefont {Lawrence~W.}\ \bibnamefont {Cheuk}},
  \bibinfo {author} {\bibfnamefont {Wenjie}\ \bibnamefont {Ji}}, \bibinfo
  {author} {\bibfnamefont {Waseem~S.}\ \bibnamefont {Bakr}}, \ and\ \bibinfo
  {author} {\bibfnamefont {Martin~W.}\ \bibnamefont {Zwierlein}},\ }\bibfield
  {title} {\enquote {\bibinfo {title} {Heavy solitons in a fermionic
  superfluid},}\ }\href {\doibase 10.1038/nature12338} {\bibfield  {journal}
  {\bibinfo  {journal} {Nature}\ }\textbf {\bibinfo {volume} {499}},\ \bibinfo
  {pages} {426--430} (\bibinfo {year} {2013})}\BibitemShut {NoStop}%
\bibitem [{\citenamefont {Ku}\ \emph {et~al.}(2014)\citenamefont {Ku},
  \citenamefont {Ji}, \citenamefont {Mukherjee}, \citenamefont
  {Guardado-Sanchez}, \citenamefont {Cheuk}, \citenamefont {Yefsah},\ and\
  \citenamefont {Zwierlein}}]{Ku2014}%
  \BibitemOpen
  \bibfield  {author} {\bibinfo {author} {\bibfnamefont {Mark J.~H.}\
  \bibnamefont {Ku}}, \bibinfo {author} {\bibfnamefont {Wenjie}\ \bibnamefont
  {Ji}}, \bibinfo {author} {\bibfnamefont {Biswaroop}\ \bibnamefont
  {Mukherjee}}, \bibinfo {author} {\bibfnamefont {Elmer}\ \bibnamefont
  {Guardado-Sanchez}}, \bibinfo {author} {\bibfnamefont {Lawrence~W.}\
  \bibnamefont {Cheuk}}, \bibinfo {author} {\bibfnamefont {Tarik}\ \bibnamefont
  {Yefsah}}, \ and\ \bibinfo {author} {\bibfnamefont {Martin~W.}\ \bibnamefont
  {Zwierlein}},\ }\bibfield  {title} {\enquote {\bibinfo {title} {Motion of a
  solitonic vortex in the bec-bcs crossover},}\ }\href {\doibase
  10.1103/PhysRevLett.113.065301} {\bibfield  {journal} {\bibinfo  {journal}
  {Phys. Rev. Lett.}\ }\textbf {\bibinfo {volume} {113}},\ \bibinfo {pages}
  {065301} (\bibinfo {year} {2014})}\BibitemShut {NoStop}%
\bibitem [{\citenamefont {Ku}\ \emph {et~al.}(2016)\citenamefont {Ku},
  \citenamefont {Mukherjee}, \citenamefont {Yefsah},\ and\ \citenamefont
  {Zwierlein}}]{Ku2016}%
  \BibitemOpen
  \bibfield  {author} {\bibinfo {author} {\bibfnamefont {Mark J.~H.}\
  \bibnamefont {Ku}}, \bibinfo {author} {\bibfnamefont {Biswaroop}\
  \bibnamefont {Mukherjee}}, \bibinfo {author} {\bibfnamefont {Tarik}\
  \bibnamefont {Yefsah}}, \ and\ \bibinfo {author} {\bibfnamefont {Martin~W.}\
  \bibnamefont {Zwierlein}},\ }\bibfield  {title} {\enquote {\bibinfo {title}
  {Cascade of solitonic excitations in a superfluid fermi gas: From planar
  solitons to vortex rings and lines},}\ }\href {\doibase
  10.1103/PhysRevLett.116.045304} {\bibfield  {journal} {\bibinfo  {journal}
  {Phys. Rev. Lett.}\ }\textbf {\bibinfo {volume} {116}},\ \bibinfo {pages}
  {045304} (\bibinfo {year} {2016})}\BibitemShut {NoStop}%
\bibitem [{\citenamefont {W.~Zwerger}(2012)}]{Zwerger2012}%
  \BibitemOpen
  \bibfield  {author} {\bibinfo {author} {\bibfnamefont {Ed.}\ \bibnamefont
  {W.~Zwerger}},\ }\href@noop {} {\emph {\bibinfo {title} {The BCS-BEC
  Crossover and the Unitary Fermi Gas}}}\ (\bibinfo  {publisher}
  {Springer-Verlag},\ \bibinfo {address} {Berlin Heidelberg},\ \bibinfo {year}
  {2012})\BibitemShut {NoStop}%
\bibitem [{\citenamefont {Machida}\ and\ \citenamefont
  {Koyama}(2005)}]{Machida2005}%
  \BibitemOpen
  \bibfield  {author} {\bibinfo {author} {\bibfnamefont {M.}~\bibnamefont
  {Machida}}\ and\ \bibinfo {author} {\bibfnamefont {T.}~\bibnamefont
  {Koyama}},\ }\bibfield  {title} {\enquote {\bibinfo {title} {{Structure of a
  Quantized Vortex near the BCS-BEC Crossover in an Atomic Fermi Gas}},}\
  }\href {\doibase 10.1103/PhysRevLett.94.140401} {\bibfield  {journal}
  {\bibinfo  {journal} {Physical Review Letters}\ }\textbf {\bibinfo {volume}
  {94}},\ \bibinfo {pages} {140401} (\bibinfo {year} {2005})}\BibitemShut
  {NoStop}%
\bibitem [{\citenamefont {Sensarma}\ \emph {et~al.}(2006)\citenamefont
  {Sensarma}, \citenamefont {Randeria},\ and\ \citenamefont
  {Ho}}]{Sensarma2006}%
  \BibitemOpen
  \bibfield  {author} {\bibinfo {author} {\bibfnamefont {Rajdeep}\ \bibnamefont
  {Sensarma}}, \bibinfo {author} {\bibfnamefont {Mohit}\ \bibnamefont
  {Randeria}}, \ and\ \bibinfo {author} {\bibfnamefont {Tin-Lun}\ \bibnamefont
  {Ho}},\ }\bibfield  {title} {\enquote {\bibinfo {title} {Vortices in
  {Superfluid} {Fermi} {Gases} through the {BEC} to {BCS} {Crossover}},}\
  }\href {\doibase 10.1103/PhysRevLett.96.090403} {\bibfield  {journal}
  {\bibinfo  {journal} {Phys. Rev. Lett.}\ }\textbf {\bibinfo {volume} {96}},\
  \bibinfo {pages} {090403} (\bibinfo {year} {2006})}\BibitemShut {NoStop}%
\bibitem [{\citenamefont {Takahashi}\ \emph {et~al.}(2006)\citenamefont
  {Takahashi}, \citenamefont {Mizushima}, \citenamefont {Ichioka},\ and\
  \citenamefont {Machida}}]{Takahashi2006}%
  \BibitemOpen
  \bibfield  {author} {\bibinfo {author} {\bibfnamefont {M.}~\bibnamefont
  {Takahashi}}, \bibinfo {author} {\bibfnamefont {T.}~\bibnamefont
  {Mizushima}}, \bibinfo {author} {\bibfnamefont {M.}~\bibnamefont {Ichioka}},
  \ and\ \bibinfo {author} {\bibfnamefont {K.}~\bibnamefont {Machida}},\
  }\bibfield  {title} {\enquote {\bibinfo {title} {{Vortex-core structure in
  neutral fermion superfluids with population imbalance}},}\ }\href {\doibase
  10.1103/PhysRevLett.97.180407} {\bibfield  {journal} {\bibinfo  {journal}
  {Physical Review Letters}\ }\textbf {\bibinfo {volume} {97}},\ \bibinfo
  {pages} {180407} (\bibinfo {year} {2006})}\BibitemShut {NoStop}%
\bibitem [{\citenamefont {Hu}\ \emph {et~al.}(2007)\citenamefont {Hu},
  \citenamefont {Liu},\ and\ \citenamefont {Drummond}}]{Hu2007}%
  \BibitemOpen
  \bibfield  {author} {\bibinfo {author} {\bibfnamefont {Hui}\ \bibnamefont
  {Hu}}, \bibinfo {author} {\bibfnamefont {Xia~Ji}\ \bibnamefont {Liu}}, \ and\
  \bibinfo {author} {\bibfnamefont {Peter~D.}\ \bibnamefont {Drummond}},\
  }\bibfield  {title} {\enquote {\bibinfo {title} {{Visualization of vortex
  bound states in polarized fermi gases at unitarity}},}\ }\href {\doibase
  10.1103/PhysRevLett.98.060406} {\bibfield  {journal} {\bibinfo  {journal}
  {Physical Review Letters}\ }\textbf {\bibinfo {volume} {98}},\ \bibinfo
  {pages} {060406} (\bibinfo {year} {2007})}\BibitemShut {NoStop}%
\bibitem [{\citenamefont {{Magierski}}\ \emph {et~al.}(2020)\citenamefont
  {{Magierski}}, \citenamefont {{Wlaz{\l}owski}}, \citenamefont {{Makowski}},\
  and\ \citenamefont {{Kobuszewski}}}]{Magierski2020}%
  \BibitemOpen
  \bibfield  {author} {\bibinfo {author} {\bibfnamefont {Piotr}\ \bibnamefont
  {{Magierski}}}, \bibinfo {author} {\bibfnamefont {Gabriel}\ \bibnamefont
  {{Wlaz{\l}owski}}}, \bibinfo {author} {\bibfnamefont {Andrzej}\ \bibnamefont
  {{Makowski}}}, \ and\ \bibinfo {author} {\bibfnamefont {Konrad}\ \bibnamefont
  {{Kobuszewski}}},\ }\bibfield  {title} {\enquote {\bibinfo {title}
  {{Spin-polarized vortices with reversed circulation}},}\ }\href@noop {}
  {\bibfield  {journal} {\bibinfo  {journal} {arXiv e-prints}\ ,\ \bibinfo
  {eid} {arXiv:2011.13021}} (\bibinfo {year} {2020})},\ \Eprint
  {http://arxiv.org/abs/2011.13021} {arXiv:2011.13021 [cond-mat.quant-gas]}
  \BibitemShut {NoStop}%
\bibitem [{\citenamefont {Inotani}\ \emph {et~al.}(2020)\citenamefont
  {Inotani}, \citenamefont {Yasui}, \citenamefont {Mizushima},\ and\
  \citenamefont {Nitta}}]{Inotani2020}%
  \BibitemOpen
  \bibfield  {author} {\bibinfo {author} {\bibfnamefont {Daisuke}\ \bibnamefont
  {Inotani}}, \bibinfo {author} {\bibfnamefont {Shigehiro}\ \bibnamefont
  {Yasui}}, \bibinfo {author} {\bibfnamefont {Takeshi}\ \bibnamefont
  {Mizushima}}, \ and\ \bibinfo {author} {\bibfnamefont {Muneto}\ \bibnamefont
  {Nitta}},\ }\bibfield  {title} {\enquote {\bibinfo {title} {{Radial
  Fulde-Ferrell-Larkin-Ovchinnikov state in a population-imbalanced Fermi
  gas}},}\ }\href@noop {} {\  (\bibinfo {year} {2020})},\ \Eprint
  {http://arxiv.org/abs/2003.03159} {arXiv:2003.03159 [cond-mat.quant-gas]}
  \BibitemShut {NoStop}%
\bibitem [{\citenamefont {Bulgac}\ and\ \citenamefont {Yu}(2002)}]{Bulgac2002}%
  \BibitemOpen
  \bibfield  {author} {\bibinfo {author} {\bibfnamefont {Aurel}\ \bibnamefont
  {Bulgac}}\ and\ \bibinfo {author} {\bibfnamefont {Yongle}\ \bibnamefont
  {Yu}},\ }\bibfield  {title} {\enquote {\bibinfo {title} {Renormalization of
  the hartree-fock-bogoliubov equations in the case of a zero range pairing
  interaction},}\ }\href {\doibase 10.1103/PhysRevLett.88.042504} {\bibfield
  {journal} {\bibinfo  {journal} {Phys. Rev. Lett.}\ }\textbf {\bibinfo
  {volume} {88}},\ \bibinfo {pages} {042504} (\bibinfo {year}
  {2002})}\BibitemShut {NoStop}%
\bibitem [{\citenamefont {Kim}\ and\ \citenamefont {Zubarev}(2004)}]{Kim2004}%
  \BibitemOpen
  \bibfield  {author} {\bibinfo {author} {\bibfnamefont {Yeong~E.}\
  \bibnamefont {Kim}}\ and\ \bibinfo {author} {\bibfnamefont {Alexander~L.}\
  \bibnamefont {Zubarev}},\ }\bibfield  {title} {\enquote {\bibinfo {title}
  {Time-dependent density-functional theory for trapped strongly interacting
  fermionic atoms},}\ }\href {\doibase 10.1103/PhysRevA.70.033612} {\bibfield
  {journal} {\bibinfo  {journal} {Phys. Rev. A}\ }\textbf {\bibinfo {volume}
  {70}},\ \bibinfo {pages} {033612} (\bibinfo {year} {2004})}\BibitemShut
  {NoStop}%
\bibitem [{\citenamefont {Salasnich}\ \emph {et~al.}(2009)\citenamefont
  {Salasnich}, \citenamefont {Ancilotto}, \citenamefont {Manini},\ and\
  \citenamefont {Toigo}}]{Salasnich2009}%
  \BibitemOpen
  \bibfield  {author} {\bibinfo {author} {\bibfnamefont {L.}~\bibnamefont
  {Salasnich}}, \bibinfo {author} {\bibfnamefont {F.}~\bibnamefont
  {Ancilotto}}, \bibinfo {author} {\bibfnamefont {N.}~\bibnamefont {Manini}}, \
  and\ \bibinfo {author} {\bibfnamefont {F.}~\bibnamefont {Toigo}},\ }\bibfield
   {title} {\enquote {\bibinfo {title} {Dc and ac josephson effects with
  superfluid fermi atoms across a feshbach resonance},}\ }\href {\doibase
  10.1134/S1054660X09040173} {\bibfield  {journal} {\bibinfo  {journal} {Laser
  Physics}\ }\textbf {\bibinfo {volume} {19}},\ \bibinfo {pages} {636--641}
  (\bibinfo {year} {2009})}\BibitemShut {NoStop}%
\bibitem [{\citenamefont {Bulgac}\ \emph {et~al.}(2014)\citenamefont {Bulgac},
  \citenamefont {Forbes}, \citenamefont {Kelley}, \citenamefont {Roche},\ and\
  \citenamefont {Wlazłowski}}]{Bulgac2014}%
  \BibitemOpen
  \bibfield  {author} {\bibinfo {author} {\bibfnamefont {Aurel}\ \bibnamefont
  {Bulgac}}, \bibinfo {author} {\bibfnamefont {Michael~McNeil}\ \bibnamefont
  {Forbes}}, \bibinfo {author} {\bibfnamefont {Michelle~M.}\ \bibnamefont
  {Kelley}}, \bibinfo {author} {\bibfnamefont {Kenneth~J.}\ \bibnamefont
  {Roche}}, \ and\ \bibinfo {author} {\bibfnamefont {Gabriel}\ \bibnamefont
  {Wlazłowski}},\ }\bibfield  {title} {\enquote {\bibinfo {title} {Quantized
  {Superfluid} {Vortex} {Rings} in the {Unitary} {Fermi} {Gas}},}\ }\href
  {\doibase 10.1103/PhysRevLett.112.025301} {\bibfield  {journal} {\bibinfo
  {journal} {Phys. Rev. Lett.}\ }\textbf {\bibinfo {volume} {112}},\ \bibinfo
  {pages} {025301} (\bibinfo {year} {2014})}\BibitemShut {NoStop}%
\bibitem [{\citenamefont {Bulgac}(2019)}]{Bulgac2019}%
  \BibitemOpen
  \bibfield  {author} {\bibinfo {author} {\bibfnamefont {Aurel}\ \bibnamefont
  {Bulgac}},\ }\bibfield  {title} {\enquote {\bibinfo {title} {Time-dependent
  density functional theory for fermionic superfluids: From cold atomic gases -
  to nuclei and neutron stars crust},}\ }\href {\doibase
  10.1002/pssb.201800592} {\bibfield  {journal} {\bibinfo  {journal} {physica
  status solidi (b)}\ }\textbf {\bibinfo {volume} {256}},\ \bibinfo {pages}
  {1800592} (\bibinfo {year} {2019})}\BibitemShut {NoStop}%
\bibitem [{\citenamefont {Wlazłowski}\ \emph {et~al.}(2018)\citenamefont
  {Wlazłowski}, \citenamefont {Sekizawa}, \citenamefont {Marchwiany},\ and\
  \citenamefont {Magierski}}]{Wlazlowski2018}%
  \BibitemOpen
  \bibfield  {author} {\bibinfo {author} {\bibfnamefont {Gabriel}\ \bibnamefont
  {Wlazłowski}}, \bibinfo {author} {\bibfnamefont {Kazuyuki}\ \bibnamefont
  {Sekizawa}}, \bibinfo {author} {\bibfnamefont {Maciej}\ \bibnamefont
  {Marchwiany}}, \ and\ \bibinfo {author} {\bibfnamefont {Piotr}\ \bibnamefont
  {Magierski}},\ }\bibfield  {title} {\enquote {\bibinfo {title} {Suppressed
  solitonic cascade in spin-imbalanced superfluid fermi gas},}\ }\href
  {\doibase 10.1103/PhysRevLett.120.253002} {\bibfield  {journal} {\bibinfo
  {journal} {Phys. Rev. Lett.}\ }\textbf {\bibinfo {volume} {120}},\ \bibinfo
  {pages} {253002} (\bibinfo {year} {2018})},\ \bibinfo {note} {arXiv:
  1711.05803}\BibitemShut {NoStop}%
\bibitem [{\citenamefont {Villois}\ \emph {et~al.}(2016)\citenamefont
  {Villois}, \citenamefont {Krstulovic}, \citenamefont {Proment},\ and\
  \citenamefont {Salman}}]{Villois2016}%
  \BibitemOpen
  \bibfield  {author} {\bibinfo {author} {\bibfnamefont {Alberto}\ \bibnamefont
  {Villois}}, \bibinfo {author} {\bibfnamefont {Giorgio}\ \bibnamefont
  {Krstulovic}}, \bibinfo {author} {\bibfnamefont {Davide}\ \bibnamefont
  {Proment}}, \ and\ \bibinfo {author} {\bibfnamefont {Hayder}\ \bibnamefont
  {Salman}},\ }\bibfield  {title} {\enquote {\bibinfo {title} {{A vortex
  filament tracking method for the Gross-Pitaevskii model of a superfluid}},}\
  }\href {\doibase 10.1088/1751-8113/49/41/415502} {\bibfield  {journal}
  {\bibinfo  {journal} {Journal of Physics A: Mathematical and Theoretical}\
  }\textbf {\bibinfo {volume} {49}},\ \bibinfo {pages} {415502} (\bibinfo
  {year} {2016})}\BibitemShut {NoStop}%
\bibitem [{\citenamefont {Villois}\ \emph {et~al.}(2020)\citenamefont
  {Villois}, \citenamefont {Proment},\ and\ \citenamefont
  {Krstulovic}}]{PhysRevLett.125.164501}%
  \BibitemOpen
  \bibfield  {author} {\bibinfo {author} {\bibfnamefont {Alberto}\ \bibnamefont
  {Villois}}, \bibinfo {author} {\bibfnamefont {Davide}\ \bibnamefont
  {Proment}}, \ and\ \bibinfo {author} {\bibfnamefont {Giorgio}\ \bibnamefont
  {Krstulovic}},\ }\bibfield  {title} {\enquote {\bibinfo {title} {Irreversible
  dynamics of vortex reconnections in quantum fluids},}\ }\href {\doibase
  10.1103/PhysRevLett.125.164501} {\bibfield  {journal} {\bibinfo  {journal}
  {Phys. Rev. Lett.}\ }\textbf {\bibinfo {volume} {125}},\ \bibinfo {pages}
  {164501} (\bibinfo {year} {2020})}\BibitemShut {NoStop}%
\bibitem{GWCode}{In the calculations W-SLDA Toolkit has been used: \url{https://wslda.fizyka.pw.edu.pl}}
\bibitem{supplemental} See supplemental online material at \{URL will be provided by the publisher \} for list of accompanying movies.
\end{thebibliography}
\end{document}